\documentclass[a4paper,fleqn,usenatbib]{mnras}

\usepackage{newtxtext,newtxmath}

\usepackage[T1]{fontenc}
\usepackage{ae,aecompl}

\usepackage{graphicx}	
\usepackage{amsmath}	
\usepackage{amssymb}	
\usepackage{pdflscape}

\newcommand{\msun}{M$_{\odot}$} 
\newcommand{\kms}{\,km\,s$^{-1}$} 
\newcommand{\ha}{H$\alpha$} 
\newcommand{\hab}{H$\alpha_{\rm b}$} 
\newcommand{\han}{H$\alpha_{\rm n}$} 

\title[Photometric detection of quiescent BHXBs]{A feasibility study on the photometric detection of quiescent black hole X-ray binaries}

\author[J. Casares and M.A.P. Torres]{Jorge Casares$^{1,2}$\thanks{E-mail: jorge.casares@iac.es}, 
Manuel A.P. Torres$^{1,2,3}$\thanks{E-mail: manuel.perez.torres@iac.es}\\
$^{1}$Instituto de Astrof\'isica de Canarias, 38205 La Laguna, Tenerife, Spain\\
$^{2}$Departamento de Astrof\'isica, Universidad de La Laguna, E-38206 La Laguna, Tenerife, Spain\\
$^3$ SRON, Netherlands Institute for Space Research, Sorbonnelaan 2, 3584 CA, Utrecht, The Netherlands \\
}
\date{Accepted XXX. Received YYY; in original form ZZZ}

\pubyear{2018}

\begin{document}
\label{firstpage}
\pagerange{\pageref{firstpage}--\pageref{lastpage}}
\maketitle

\begin{abstract}
We investigate the feasibility of detecting quiescent black hole X-ray binaries using optical 
photometric techniques. To test this we employ a combination of $r$-band and \ha~filters currently available 
at the Roque de los Muchachos Observatory. Photometric observations of four dynamical black holes (GRO 
J0422+320, A 0620-00, XTE J1118+480 and XTE J1859+226) at SNR$\gtrsim$35-50, supplemented with near 
simultaneous spectroscopic 
data, demonstrate that it is possible to recover the FWHM of the \ha~emission line to better than 10\%~for targets 
with a wide range of line EWs and down to magnitude r$\sim$22. 
We further explore the potential of our photometric system to disentangle other populations of  compact stars and 
\ha~emitters. In particular, we show that {\it HAWKs}, a survey designed to unveil quiescent black holes, will also provide 
a detailed census of other Galactic populations, most notably short period (eclipsing) cataclysmic 
variables, neutron star X-ray binaries and ultra-compact binaries.

\end{abstract}




\section{Introduction}
\label{intro}

In the era of gravitational wave (GW) discoveries \citep{abbott16a, abbott16b, abbott17a, abbott17b, abbott17c} 
the study and characterisation of accreting black holes (BH) in the Milky Way remains a topic of 
important strategic interest. 
These systems provide us with a reference sample of BH properties (e.g. space density, masses, spin) 
stemmed from well defined evolutionary channels at high metallicity (see \citealt{tauris06}). 
And yet, our knowledge on the formation of black hole X-ray binaries (BHXBs) is far from complete, 
with crucial issues that need to be solved. Just to mention a few, it is 
not well understood how a low mass companion star can possibly survive the common envelope phase and a 
supernova (SN) explosion, resulting in the observed numbers of BHXBs (e.g. \citealt{podsiadlowski03,wang16}). 
It is uncertain whether BHs receive a natal kick or are formed by implosion \citep{repetto17,mirabel17,casares17}.  
Is also unclear if standard BHXB formation channels can produce BHs heavier than $\sim$15 
\msun~ (c.f. \citealt{belczynski10}) or whether SN physics is responsible for the $\sim$2-5 \msun~mass gap between 
neutron stars (NS) and BHs hinted by observations \citep{fryer12,ugliano12}. 

With the exception of Cyg X-1 and MWC 656 (a Be/BH binary that may end up in a BH/NS merger and thus a 
source of GWs; \citealt{casares14, grudzinska15}) the great majority of accreting BHs in the Galaxy have been 
detected through dramatic X-ray outbursts. About 60 of these, so-called,  BH {\it X-ray transients} have been 
discovered in five decades of X-ray surveys (see the BlackCat catalogue in \citealt{corral16}) but only 17 hold  
dynamical confirmation (i.e. mass function greater than $\sim$3 \msun), owing to difficulties in measuring the 
spectrum of the companion star at very faint quiescent luminosities \citep{casares-jonker14}. 
Our knowledge of their fundamental parameters (orbital period, masses, space velocities, etc.), and thus on the formation 
and evolution of BHXBs as a population, is clearly jeopardised by limited statistics. Therefore, it is of paramount 
interest to explore new routes to unveil the hidden population of hibernating (quiescent) BHXBs.  

Notwithstanding these limitations, dynamical information can still be extracted from scaling relations based on the 
properties of the disc \ha~emission line (\citealt{casares15, casares16}; Papers I and II hereafter). In particular, 
building  upon the FWHM-$K_2$ empirical relation, presented in Paper I, we have developed a new approach to  
single out quiescent BHXBs among the myriad of \ha~emitters. 
In fact, blind \ha~surveys of the Galactic plane, such as IPHAS, the {\it INT Photometric  
\ha~Survey of the Northern Galactic Plane} \citep{drew05}) have successfully increased the statistics of  
\ha~emitting populations, including young stellar objects (YSOs), cataclysmic variables (CVs), symbiotic stars and 
others, but have so far failed to discover quiescent BHXBs. This is unsurprising given the extremely low density of 
the latter and the lack of clear optical signatures that set them apart from other populations of \ha~emitters.   
A different strategy, the selection of \ha~sources  
with weak X-ray emission from Chandra surveys of the Galactic Bulge and Plane \citep{grindlay05,jonker11} 
has proved most sensitive to magnetic CVs and coronal stars but, again, not to quiescent BHXBs 
(see e.g. \citealt{rogel06,torres14,wevers17}). 

Alternatively, in \cite{casares18} (hereafter Paper III), we propose the 
full-widh-half-maximum (FWHM) of the \ha~line as an efficient diagnostic to 
discriminate BHXBs from other \ha~emitting objects. 
Paper III presents a proof-of-concept on how \ha~widths can be extracted from imaging techniques and devises a new 
photometric system, optimised to measure equivalent widths (EWs) and FWHMs, the two basic line-profile parameters. 
It is based on three \ha~filters with squared response functions of increasing width but the same central wavelength. 
This allows breaking the degeneracy between EW and reddening \citep{drew05, witham06}, and thus 
a unique determination of both EW and FWHM line values. 
Furthermore, a filter cycling strategy is proposed to mitigate the effect of flickering variability in FWHM determinations while 
$\sim$1 kilo square degree survey ($HAWKs$, the {\it H$\alpha$ Width Kilo-degree survey}) at signal-to-noise ratio 
(SNR) $\sim$50 and depth $r=22$ is set out for the discovery 
of, at least, $\sim50$ new hibernating BHXBs. Here in this paper we present a feasibility study to demonstrate that this strategy 
allows the recovery of FWHM values in quiescent BHXBs to better than 10\% accuracy, through photometric observations of a 
sample of quiescent BHXBs (Section~\ref{sec:analysis}).  The sample embraces BHXBs with a large range of EW and 
FWHM values, down to very faint magnitudes $r\sim$22. In Section~\ref{sec:summary} we 
summarize the results and lay out the prospects of this photometric system for isolating other populations of 
compact stars and \ha~emitters.  

\section{Observations and data reduction}
\label{sec:obs}

\begin{figure}
	\includegraphics[angle=0,width=\columnwidth]{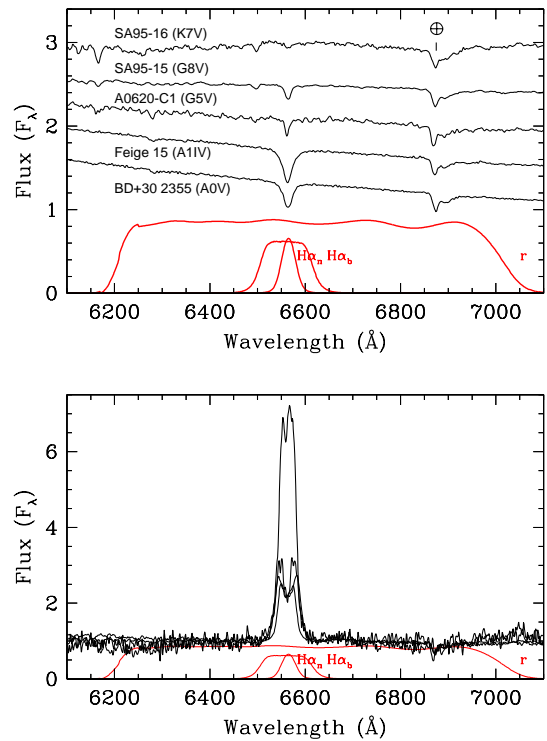}
    \caption{Top: WHT+ACAM spectra of standard stars, together with the transmission 
    curves of the r-band and the two \ha~filters. The GTC+OSIRIS spectrum of a comparison star in the field of 
    A 0620-00, broadened to the same resolution of the ACAM spectra, 
    is also displayed. All the spectra have been normalized to the continuum at \ha. For the sake of display, relative vertical 
    offsets have been applied to the spectra.  
    Bottom: Averaged GTC+OSIRIS spectra of the four BHXBs. These spectra have also been normalized to the continuum 
    at \ha.}
    \label{fig:fig1}
\end{figure}

We have employed the Auxiliary-port CAMera (ACAM) on the 4.2 m William Herschel Telescope (WHT) at the 
Roque de los Muchachos Observatory in La Palma to 
obtain images of five quiescent BHXBs: Swift J1357-0933 on the night of 16 Feb 2018, GRO J0422+320, A 0620-00 and 
XTE J1118+480 on 17 Feb 2018 
and XTE J1859+226 on 20 June 2018. We name these targets J1357, J0422, A0620, J1118 and J1859 hereafter. 
The images were obtained with the NOT29 broad \ha~filter ($\lambda_{\rm central}$=6560 
\AA, FWHM=113 \AA), the NOT21 narrow \ha~filter ($\lambda_{\rm central}$=6564 \AA, FWHM=33 \AA) and the r-band 
filter MR661 ($\lambda_{\rm central}$=6608 \AA, FWHM=798 \AA) from OASIS, a former Isaac Newton Group  (ING) 
instrument currently decommissioned. The latter has been chosen among a possible list of ING r-band filters 
because it has the closest effective wavelength to the \ha~rest wavelength, a critical requirement of 
our photometric system (see Paper III for details). The filters are hereafter referred to as \hab, \han~and r and their 
transmission profiles\footnote{The transmission curves are available from 
http://www.not.iac.es/instruments/filters/curves-ascii/29.txt, 
http://www.not.iac.es/instruments/filters/curves-ascii/21.txt and 
http://catserver.ing.iac.es/filter/filtercurve.php?format=txt\&filter=585} are plotted in Fig.~\ref{fig:fig1}. 

The r-band filter has a small clear aperture of 25mm and it was mounted in the ACAM slit unit, located at 
the focal-plane of the instrument. By doing this we ensure that the filter will not vignette the light beam although the available 
field-of-view (FOV) becomes severely limited, with only the central 1.1 arcmin (diameter) unvignetted. 
The two \ha~filters were 
instead mounted in filter wheel positions, at the pupil-plane of the instrument. This introduces a blueshift in the effective 
wavelength with distance from the optical axis, but the effect is negligible within the central 1.1 arcmin area of overlap 
between our three filters i.e. the effective FOV for useful scientific observations. The 2Kx4K EEV CCD was windowed to 
the central part (of approximately 1.6 arcmin side)  resulting in a readout time of only two seconds. 

Continuous r/\hab/\han~cycles were performed on every BH target. The number of 
cycles and integration times were initially designed to reach a final (average) SNR$\gtrsim$50 in every filter, a requirement 
defined by Paper III. Four such cycles were obtained for J0422 between 20:17-20:57 UT, five for A0620 between 
21:05-21:18 UT, 16 cycles for J1118 between 22:16-23:30 UT and four cycles for J1859 between 02:03-03:10 UT. 
The J1357 observations were performed just before morning twilight and consisted of a single r/\hab/\han~cycle. 
The nights of 16 and 17 Feb were clear and photometric, with seeing 
around 1 arcsec, except for the block of J1118 observations when seeing degraded to $\sim$2.5 arcsec, with rapid 
variations caused by wind gust conditions. The night of 20 June was also photometric, with excellent seeing of 0.6 
arcsec.

Near simultaneous spectroscopic observations of J0422, A0620, J1118 and J1859 were programmed with the 
Optical System 
for Imaging and low-Intermediate-Resolution Integrated Spectroscopy (OSIRIS) at the 10.4m Gran Telescopio Canarias 
(GTC). We employed grism R1000B and a 1.0 arcsec slit to cover the wavelength range 3780-7880 \AA~at 6.4 
\AA~resolution (=292 \kms~at \ha). A spectrum of the flux standard BD+52 913 was also acquired with a slit width of 
2.5 arcsec for the purpose of flux calibration. The slit was oriented at parallactic angle to minimize the impact of 
atmospheric refraction on our flux calibration. The J0422 spectroscopy spans over 95\% of the corresponding 
photometric window while the A0620 and J1859 spectra cover 85\% and 74\% of their photometric baselines, 
respectively. A technical 
failure during the GTC observations of J1118 produced a 41 min gap with no useful data which results in only 
45\% simultaneous coverage. The J1357 photometry could not be supported by simultaneous GTC spectroscopy. 

r/\hab/\han~images of two late-type photometric Landolt stars (SA95 15 \&16, \citealt{landolt92}) and the A0V star 
BD+30 2355 were also obtained on the night of 17 Feb. The latter was acquired to provide a zero point in the photometric 
calibration tied to the Vega system. Low resolution spectra of the photometric standards were further acquired with the 
V400 grism available on the filter-wheel unit of ACAM. These ACAM spectra cover the wavelength range 5020-9280 \AA~at  
$\sim$495 \kms~resolution.  The aim was to compute synthetic magnitudes with the nominal transmission 
curves of our filters, to be compared to the real magnitudes provided by the actual filters. 
An ACAM spectrum of the spectroscopic flux standard Feige 15 (plus a set of r/\hab/\han~images) was also obtained 
for the purpose of flux calibrating the spectra of the photometric standards.  
Our list of standards also includes a star in the field of A0620, USNO B1.0 0896-0086799, 
which fortuitously lay in the OSIRIS slit. We name this star A0620-C1 hereafter. 
An observing log, with details on integration times, is presented in Table~\ref{tab:tab1}. 

The spectroscopic data were processed in the standard way with debias, flat-field correction and optimal spectral 
extraction using STARLINK/PAMELA routines \citep{marsh89}. Observations of CuNe+CuAr (ACAM) and 
HgAr+Ne (OSIRIS) lamps were employed to derive a pixel-to-wavelength calibration through a 4th order polynomial 
fit to more than 28 lines across the entire wavelength range. Small flexure corrections, obtained from the position of 
the \ion{O}{I} 5577.34 and 6300.30 sky lines, were applied to individual spectra in order to match the laboratory 
positions within 1 \kms. Figure~\ref{fig:fig1} displays the OSIRIS (average) spectra of the BHXBs and the ACAM 
spectra of the standard stars,  together with the transmission curves of our filters. We have assigned an approximate 
spectral classification for the Landolt photometric standards based on the relative depth of the spectral lines and 
their photometric (B-V) colour.   

The photometric data acquired with ACAM were reduced in the following way: for each program object and standard star,  
images were bias subtracted, flat-field corrected and aligned using 
{\small{{\texttt{IRAF}}}}\footnote{{\small{{\texttt{IRAF}}}}  is distributed by the National Optical Astronomy Observatories, which 
are operated by AURA, Inc., under cooperative agreement with
the National Science Foundation.} tasks. 
Stellar fluxes were extracted using aperture photometry because none of our targets is blended with nearby stars.
Aperture photometry was then performed on each image  using the 
{\small\texttt{DAOPHOT}} package to derive instrumental magnitudes for our targets, their field stars as well as  the 
Landolt standards. Different apertures were chosen  according to the image quality at each filter. Aperture corrections 
were subsequently calculated with a curve-of-growth analysis using {\small\texttt{DAGROW}} \citep{stetson90} and 
applied to the instrumental magnitudes.

\section{Analysis}
\label{sec:analysis}

Following usual convention, we decide to refer our magnitudes to the Vega based zero magnitude scale.  
For example, the magnitude of a given target in the \hab~filter will be provided by 

\begin{equation}
m_{H\alpha_b}= -2.5 \log \left( \frac{F_{H\alpha_b}}{F^{0}_{H\alpha_b}} \right)
 \label{eq:mag}
\end{equation}

\noindent
where $F_{H\alpha_b}$ is the target's flux while $F^{0}_{H\alpha_b}$ that of the A0V star BD+30 2355. 
Note that, by adopting this definition, the magnitudes (and colours) of a A0V standard star are set to be zero. 
As a sanity test, we start by comparing the observed magnitudes of the standard stars with their synthetic magnitudes. 
The latter are obtained from equation~\ref{eq:mag}, using simulated fluxes computed through the  
convolution of the filter's response ($T_{H\alpha_b}$) with the observed spectra i.e. 

\begin{equation}
F_{H\alpha_b}= \int{T_{H\alpha_b}\times f_{\lambda}~d\lambda}
 \label{eq:convolution1}
\end{equation}

\noindent
and 

\begin{equation}
F^{0}_{H\alpha_b}= \int{T_{H\alpha_b}\times f^{0}_{\lambda}~d\lambda}
 \label{eq:convolution2}
\end{equation}

\noindent
where $f_{\lambda}$ and $f^{0}_{\lambda}$ are the spectra of the target and BD+30 2355, respectively. In all cases, 
both the observed spectra and filter response curves have been re-sampled to a common bin size of 1 \AA~pix$^{-1}$. 
Table~\ref{tab:tab2} lists the synthetic and observed photometric colours of the four standard stars plus A0620-C1. 
The mean differences between photometric and synthetic colours $\left(m_{r}-m_{H\alpha_b}\right)$ and 
$\left(m_{H\alpha_b}-m_{H\alpha_n}\right)$ are 0.03$\pm$0.05 and 0.02$\pm$0.03 respectively. These 
values drop to 0.01$\pm$0.03 and -0.001$\pm$0.019 if we restrict ourselves to the three late-type 
standards. This indicates that the A-type stars are responsible for most of the difference, probably 
caused by the ACAM spectra not properly resolving the core of the broad \ha~absorption profiles. In any case we 
consider these deviations acceptable, given our limitations in spectral resolution and uncertainties in flux calibration. 
It should be noted that the photometric $m_r$ values listed in the second column of Table~\ref{tab:tab2} are purely 
instrumental. Comparison with calibrated r-band magnitudes indicates that the former are underestimated by $\simeq$1.1 mag.  

\subsection{Calibration of Photometric EWs and FWHMs}
\label{sec:calibration}

As explained in appendixes B and C of Paper III, for the case of perfect ideal filters 
(i.e. those with squared response curves with 100\% peak transmission and identical central wavelength) 
it is possible to derive the EW and FWHM of an \ha~line through equations

\begin{equation}
EW^{\star}=\frac{W_{r} \times \left(\frac{F_{H\alpha_b}}{F_r}\right)- W_{H\alpha_b}}{1-\left(\frac{F_{H\alpha_b}}{F_r}\right)} 
\label{eq:pew}
\end{equation}

\begin{equation}
FWHM^{\star}=\frac{EW^{\star}}{\left(\frac{EW^{\star}+W_{H\alpha_b}}{W_{H\alpha_n}}\right)\times\left(\frac{F_{H\alpha_n}}{F_{H\alpha_b}} \right)-1} 
\label{eq:pw}
\end{equation}

\noindent
where $EW^{\star}$ and $FWHM^{\star}$ are the EW and FWHM values, as measured from the filters, 
$W_{r}$, $W_{H\alpha_b}$, $W_{H\alpha_n}$ the equivalent widths 
of the r-band, \hab~and \han~filters and $F_{H\alpha_b}$, $F_{H\alpha_b}$, $F_{H\alpha_n}$ the associated fluxes. 
For the case of real filters, however, variations in filter response curves and central wavelength relative to those of ideal filters 
will introduce deviations between the $EW^{\star}$ and $FWHM^{\star}$ measurements and the true line EW and FWHM 
values. For the simulated filters presented in Paper III, we showed that simple calibration constants  can 
account for the observed deviations (eqs. B4 and C3 in Paper III). 
Instead, we find here that the MR661, NOT29 and NOT21 filters do require a quadratic calibration term i.e. 

\begin{equation}
EW_{ph}=EW^{\star} \left(C_{1}+C_{0}~EW^{\star}\right)
\label{eq:pew_cal}
\end{equation}

 \begin{equation}
FWHM_{ph}=FWHM^{\star} \left(C_{2}+C_{3}~FWHM^{\star}\right)  
\label{eq:pw_cal}
\end{equation}

\noindent
where $C_{0}-C_{3}$ are the calibration constants and $EW_{ph}$ and $FWHM_{ph}$ our best photometric 
determination of the line EW and FWHM. 
We note that higher order polynomial terms do not lead to a significant improvement in the calibration of the 
photometric parameters. 
As in Paper III, we derive the calibration constants 
by comparing EW and FWHM values from a grid of simulated double-peaked emission \ha~profiles with synthetic 
values obtained through convolution with the filter transmission curves and equations~\ref{eq:pew}-\ref{eq:pw_cal}, 
with $W_{r}$=674 \AA, $W_{H\alpha_b}$=73 \AA~and $W_{H\alpha_n}$=25 \AA.
This results in $C_{0}=0.0009$, $C_{1}=1.8588$, $C_{2}=0.4200$ and $C_{3}=0.0003$. 
 
 \begin{figure}
	\includegraphics[angle=0,width=\columnwidth]{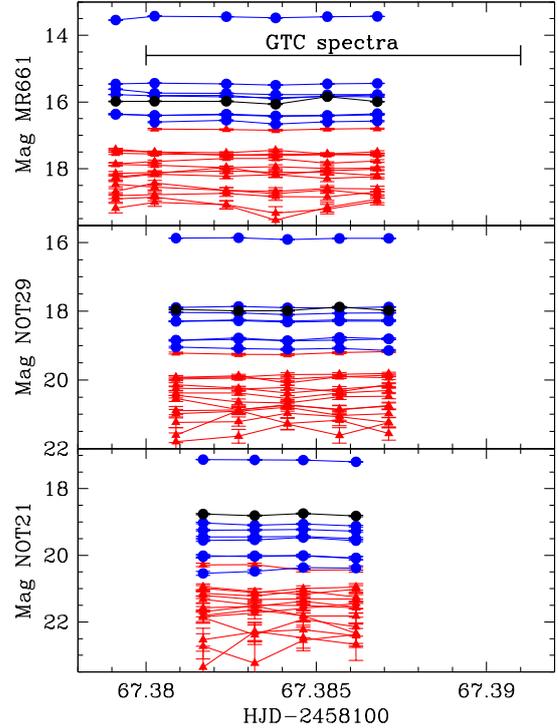}
    \caption{Light curves of A 0620-00 (black solid circles) and 22 stars in the 1.1 arcmin field of view. Displayed 
    magnitudes are instrumental. As a reference, calibrated r-band magnitudes correspond to mag(MR661)+1.1.    
Blue solid circles indicate field stars with SNR$\ge$50 in the three filters, the remaining are marked by red solid triangles 
and no longer considered.  Three field stars lie in the outer vignetted area of the first (offset) image of the MR661 filter 
and, therefore, are not displayed in the top panel. The last NOT21 image is corrupted by CCD cross-talk and has 
been rejected. 85\% of the photometric window is covered by GTC spectroscopy.} 
    \label{fig:fig2}
\end{figure}

 \subsection{Photometric EWs and FWHMs}
\label{sec:photom} 

We can now apply the calibrated relations eq.~\ref{eq:pew}-\ref{eq:pw_cal} to our ACAM photometry and derive 
$EW_{ph}$ and $FWHM_{ph}$ values for each BHXB. 
As explained in Section~\ref{sec:obs}, the photometric data were acquired as a sequence of consecutive 
r/\hab/\han~cycles so to average out flickering variability, typical of quiescent BHXBs \citep{zurita03}).  
This leads to a set of quasi-simultaneous light curves for each object in the three filters. An example is presented  
in Figure~\ref{fig:fig2}, where the light curves of A0620 and 22 field stars are displayed. 
Stars with SNR$<$50 in any filter are rejected by a clipping process. 
Time averaged magnitudes in every filter were subsequently computed as the weighted mean of  
individual data points, and the resulting photometric colours of BHXBs transformed into flux ratios and introduced into 
equations~\ref{eq:pew}-\ref{eq:pw_cal} to derive $EW_{ph}$ and $FWHM_{ph}$ values.

\begin{figure}
	\includegraphics[angle=0,width=\columnwidth]{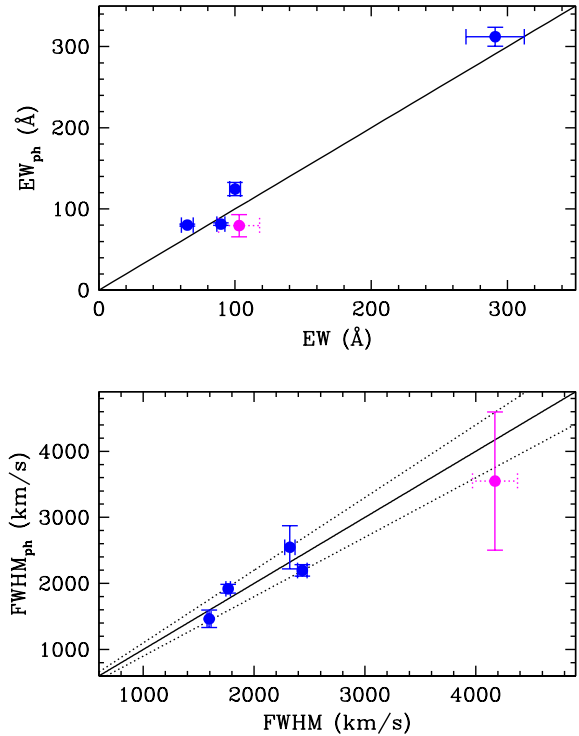}
    \caption{Top: Photometric EWs ($EW_{ph}$) versus EW values measured from near-simultaneous spectra. 
    Bottom: Photometric FWHMs ($FWHM_{ph}$) versus FWHM values from spectra. The magenta points represent 
    values for J1357 derived from a single photometric point (SNR$\sim$16 per filter). In this case, due to the lack of simultaneous 
    spectroscopy, we adopt EW and FWHM values from Mata Sanchez et al. (2015). 
    The dotted lines represent the goal of $\pm$10 \% fractional  limit in $FWHM_{ph}$.} 
    \label{fig:fig3}
\end{figure}

To assess how reliable these photometric measurements are we have also extracted EW and FWHM values from the 
near-simultaneous GTC spectra. EWs were obtained by integrating the \ha~flux in the (continuum normalized) 
spectra, while FWHMs came from single Gaussian fits to the \ha~profiles. Following Paper I, we adopt the mean  
and standard deviation in the distribution of individual values. 
A comparison between the photometric and spectroscopic determinations of EW and FWHM values 
is presented in Figure~\ref{fig:fig3} and Table~\ref{tab:tab3}. 
The first two columns of Table~\ref{tab:tab3} provide additional information on the accumulated $m_r$ magnitudes and SNR. 
We stress again the fact that the quoted $m_r$ magnitudes are instrumental. Calibrated $r$-band (continuum) magnitudes 
are given in the third column 
and have been estimated through 
$r=m_{r} + 1.1+2.5 \log \left( 1+ EW/W_{r} \right)$, 
where the latter term accounts for the contribution of the $H\alpha$ flux to the  $m_r$ magnitude.  
For completeness, Table~\ref{tab:tab3} and Figure~\ref{fig:fig3} also include the $EW_{ph}$ and $FWHM_{ph}$ 
determinations for J1357, as obtained 
from a single photometric observation. Due to the lack of simultaneous GTC spectroscopy, however,  we here adopt 
spectroscopic values from spectra obtained in 2014 by \cite{mata15}. These values are fully consistent with 
an earlier determination reported in \cite{torres15}. 

\begin{figure}
	\includegraphics[angle=-90,width=\columnwidth]{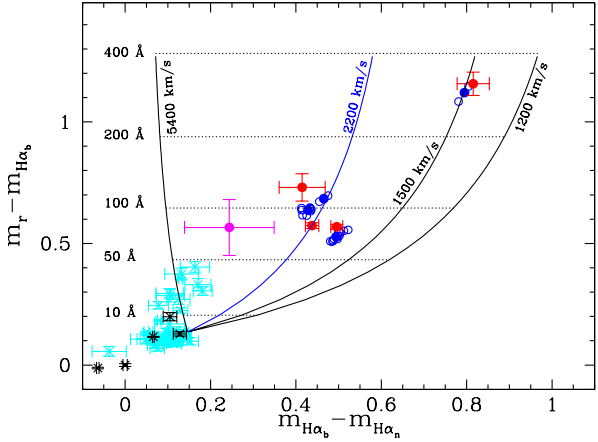}
    \caption{The \ha~colour-colour diagram. The photometric colours of standard stars are represented by black asterisks, with the 
    zero point defined by the A0V star BD +30 2355. Average photometric colours of BHs and 
    field stars with SNR$\ge$50 are indicated by filled red circles and cyan crosses, respectively. For comparison, synthetic colours 
    of individual GTC BH spectra are plotted as blue open circles, while blue filled circles represent synthetic colours from average 
    spectra. The solid magenta circle depicts the photometric colours of Swift J1357-0933, as derived from a single set of images. 
    Lines of constant EW, in the range 10-400 \AA, and constant FWHM between 1200-5400 \kms~are also marked.
} 
    \label{fig:fig4}
\end{figure}

Overall we observe a good agreement between FWHM and $FWHM_{ph}$ values, 
with $<$10\% fractional difference, which was our initial test goal. 
Only in the case of J1357 does the difference rise to 15\%, although the large error bar in 
$FWHM_{ph}$ makes the two values consistent within 1 $\sigma$. 
In any case, it should be borne in mind that J1357 experienced an outburst in April 2017 \citep{drake17} 
and, therefore, the binary may have not returned to the pre-outburst quiescent state by the time of our ACAM 
observations. In that case, the $FWHM_{ph}$ value would be underestimated because the accretion disc 
may not have time to reach the equilibrium radius 
(see e.g. Fig. 2 in Paper I for the long-term evolution of FWHM in V404 Cyg).  Following 
from Paper III, we also present the $EW_{ph}$ and $FWHM_{ph}$ information in the form of a 
colour-colour diagram\footnote{We note in passing a small error in Fig. 8 of Paper III, which appears flipped across the 
x-axis according to its label.} in Figure~\ref{fig:fig4}. To guide the eye, lines of constant EW and FWHM, as computed 
from eq.~\ref{eq:pew}-\ref{eq:pw_cal}, are overplotted.  

As a further test, we have calculated synthetic magnitudes through the convolution of the BHXB spectra with the filter 
transmission curves. The results are again listed in Table~\ref{tab:tab3} and plotted as blue circles in 
Figure~\ref{fig:fig4}, with open circles referring to synthetic magnitudes of individual spectra while filled circles to those 
of average spectra. 
We observe that line profile variability causes slight changes in the position of the blue open circles across the diagram. 
Since these fluctuations are sampled on short timescales ($\sim$1-30 min), they are likely dominated by stochastic flickering 
rather than smooth orbital modulations. Interestingly, the displacements appear to follow lines of constant FWHM, 
suggesting that flickering is dominated by EW variations rather than changes in line width. This agrees with 
previous spectrophotometric studies of V404 Cyg where it was found that flaring activity is better traced by line flux  
than continuum flux, with the width and shape of the line profile remaining largely unaffected \citep{hynes02,hynes04}. 

The synthetic FWHM values ($FWHM_{syn}$) presented in Table~\ref{tab:tab3} are seen to differ from 
$FWHM_{ph}$ by typically $\sim$3-12\%. On the other hand, $FWHM_{syn}$ deviates by 3-6\%~from the spectroscopic 
FWHM values. Since these are obtained from the same data but through different 
methods (synthetic photometry versus direct spectral line fitting), the latter discrepancies reflect our limitation in 
calibrating the photometric quantities $EW_{ph}$ and $FWHM_{ph}$ 
(Section~\ref{sec:calibration}). The result is not surprising, given the different transmission curves 
and central wavelengths of our actual filters compared to those of perfect ideal filters.   

\section{Summary and Outlook}
\label{sec:summary}

The outcome of our feasibility study is summed up by Table~\ref{tab:tab3} and 
Figures~\ref{fig:fig3}-\ref{fig:fig4}. The conclusions are summarised as follows: 

\begin{itemize}
\item[(1)]  \ha~line widths can be measured to about $\sim$10\%~accuracy through images obtained with 
an adequate combination of filters. This becomes possible even at very faint luminosities,  
provided  SNR$\approx$50 is achieved in every filter. For example, in the case of J0422, with $r=21.3$,  
SNR$\sim$50 was acquired in all filters, leading to 
an 8\%~fractional difference between $FWHM_{ph}$ and the (spectroscopic) FWHM. 

\item[(2)] Line width precision significantly better than $\sim$10\%~is not always possible because of intrinsic 
fluctuations, mostly driven by flickering variability. As an example, very high SNR in the range $\sim$150-200 
was acquired for the bright targets A0620 and J1118 but this only results in $FWHM_{ph}$ determinations  
with 9-10\%~fractional difference with respect to the spectroscopic FWHM values.  

\item[(3)]  As observed in Figure~\ref{fig:fig4}, flickering variability leads to EW fluctuations on $\sim$min 
timescales, with little variation in line FWHM. In any case, we have shown that these can be averaged out 
using a filter cycling strategy. This, in turn, allows considerable extension of the dynamic range and prevents 
saturation of relatively bright stars. For instance, our deep J0422 observation achieves an accumulated 
SNR$\sim$50 at $r\sim21$ while millimag precision is obtained for field stars with $r\sim16$ . 
Furthermore, filter cycling provides light curve information (Figure~\ref{fig:fig2}) which becomes extremely 
useful to identify short-period CVs through the presence of $\sim$2-3 mag deep eclipses.  
As stated in Paper III, these are the main Galactic sources of BHXB contamination at very large FWHM values 
$\ge$2200 \kms.

\item[(4)] In the case of the faintest target J1859 ($r\simeq$22) our observations prove that BHXBs can still be recovered 
with SNR$\sim$35. 
This is equivalent to the SNR expected at $r$=23 for a survey goal of SNR=50 at $r$=22. 
In other words, the J1859 observation demonstrates   
that BHXBs with FWHM$\geq$2200 \kms~can be detected with 3\% photometry at $r$=23. 
According to Paper III, such survey depth would lead to $>$100 new BHXBs 
in an area of 1 kilo square deg on the Galactic Plane, an order of magnitude improvement over the 
current population.    

\item[(5)] The single photometric observation of J1357 also 
demonstrates that BHXBs with very large FWHM$\sim$3000-4000 \kms~can be identified above the 2200 \kms~limit 
even at modest SNR$\sim$16. For a survey goal of SNR=50 at $r$=22 this implies that J1357-like binaries can be detected 
down to $r\sim$24.5. As shown by Figure 9 in Paper III, these are all short period BHXBs and might  
represent the bulk of the hidden hibernating population. 
It should be noted that, because line width increases at short orbital periods, our FWHM selection method is actually 
biased towards detecting short period quiescent BHs. This is opposed to X-ray/radio survey efforts that are biased 
to selecting long period BHs since luminosity decreases with period \citep{wu10,knevitt14}.  
Furthermore, since FWHM increases with binary inclination as well (eq. 8 in Paper I) our strategy is also biased towards detecting 
high inclination BHs. This makes another interesting outcome, given the current paucity of BHXBs with i$\gtrsim75^{\circ}$  due to  
X-ray selection effects \citep{narayan05}.
  
\item[(6)]  Finally, Figure~\ref{fig:fig4} shows that field stars appear conveniently segregated from BHXB targets. 
G and early-K type stars tend to cluster at EW$\simeq$0, near the focus point where lines of equal FWHM 
converge. Earlier spectral types lie along a tail leftwards of the focal point while late-K and M stars define a 
vertical stream along FWHM$\approx$5400 \kms, driven by the appearance of TiO molecular bands on each side 
of \ha, which effectively fake extremely broad emission profiles.  

\end{itemize}

An important asset of our filter combination 
is the clean separation of \ha~emitters from field 
stars, independently of interstellar reddening. This is possible because all filters are centered at the rest 
wavelength of \ha. In fact, a relative displacement between the central wavelength (CWL) of the $r$-band 
and \hab~filters would lead to a vertical shift in the position of reddened field stars, while a displacement between 
the CWLs of the \hab~and \han~filters would result in a horizontal shift. 
We have quantified this by running simulations using idealized nearly-squared filters with effective widths 37 \AA, 150 \AA~and 
350 \AA, and the Jacoby library of standard stars \citep{jacoby84}, reddened by several amounts. 
We here adopt \han~and \hab~filters that are broader than NOT21 and NOT29, motivated by the scientific 
requirements presented Paper III. We also limit the FWHM of the $r$-band filter to 350 \AA~in order to avoid 
the main telluric bands and strong airglow emission lines such as \ion{O}{I} 6300 \AA. We find that, for  
extreme reddenings E(B-V)=3, a displacement of +30 \AA~(-30 \AA) between the CWL of \hab~and that 
of the $r$-band translates into a vertical shift of +0.05 (-0.05) mags in the diagram. Similarly, a horizontal shift of 
+0.05 (-0.05) mags is obtained if the CWL of \han~is displaced by +30 \AA~(-30 \AA) with respect to that of the \hab~filter. 
These shifts are very modest and imply that BH candidates (and other \ha~emitters), detected by \ha~filters 
with $\pm$15 \AA~tolerance in CWL, will not be mixed up with countless field stars, even along sight-lines 
of substantial interstellar extinction.  

To conclude the paper we now focus our attention on opportunities presented by our photometric system to 
disentangle other populations of compact stars and \ha~emitters.  

 \subsection{Other Galactic Populations in HAWKs}
\label{sec:hawks} 

\begin{figure}
	\includegraphics[angle=0,width=\columnwidth]{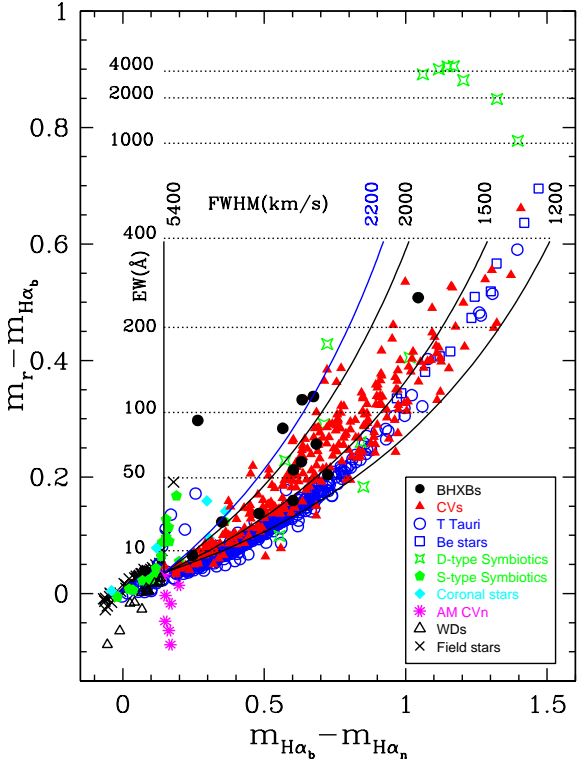}
    \caption{\ha~colour-colour diagram showing synthetic colours of BHXBs and field stars, together with other Galactic populations of interest. The list of BHXBs contains spectra from this paper together with others presented in Paper I and Mata S\'anchez et al. (2015). Other  spectral samples have been selected from the SDSS and IPHAS catalogues. This figure has been produced using a set of idealised nearly-squared filters centered at 6563 \AA~and effective widths 37 \AA, 150 \AA~and 350 \AA.  
Lines of constant EW and FWHM are represented as in Figure~\ref{fig:fig4}. 
     } 
    \label{fig:fig5}
\end{figure}

$HAWKs$, a survey concept based on these filters, will  not only boost the statistics of hibernating BHs. 
It will also deliver a full census of \ha~emitters and other Galactic populations to unprecedented depths. 
$HAWKs$ will effectively unfold into a catalogue of dynamical BHs (the $BLACK-HAWKs$) plus other 
catalogues of different ``flavours'' (the $``COLOURED''-HAWKs$), broadly classified by their FWHM-EW 
positions in the \ha~colour-colour diagram. This is illustrated in Figure~\ref{fig:fig5}, where 
synthetic colours of several Galactic populations of interest are represented  
(note that this figure has been produced using the same idealised filters referred to above rather than the actual 
MR661, NOT21 and NOT21 filters employed in Figure~\ref{fig:fig4}).

For example, $HAWKs$ will produce a new census of symbiotic stars i.e. long period accreting binaries with a compact 
star embedded in the wind of an evolved giant companion \citep{belczynski00}. 
D-type symbiotics tend to cluster at very large EWs $>$1000 \AA, a region also populated by 
PNe. S-type symbiotics, on the other hand, are dominated by the spectrum of the donor and distribute along  
the main stellar locus defined by field stars. 
The size of the symbiotic population is very uncertain ($\approx10^{3}-10^{5}$) with only $\sim$200 binaries 
currently known, 19 of which have been revealed by IPHAS \citep{rodriguez-flores14}. 
$HAWKs$, with its deeper survey limit, can make a large impact on the field.  

Young stellar objects (YSOs) of different types, such as T Tauri, Herbig Haro objects or 
classical Be stars, are characterized by narrow FWHMs in the range $\sim$100-500 \kms~ 
and, thus, will concentrate at the lower FWHM limit of the diagram. 
A much expanded sample will allow important questions on YSO physics, such as the 
evolution of accretion rates and the survival of protoplanetary discs, to be addressed \citep{barentsen11,venuti18}. 

On the other hand, CVs do spread over a much wider range of FWHM values. Those with 
FWHM$\ge$2200 \kms~will reveal themselves as short-period eclipsing WZ Sge and minimum period-bouncers, 
of considerable interest for the study of CV evolution (e.g. \citealt{littlefair08, gansicke09}). CVs with  
FWHM in the range 1500-2200 \kms~will be mostly eclipsing too, allowing for precise white dwarf mass 
determinations and the identification of possible SN Ia progenitors \citep{maoz14}.  
Within this parameter space, the sample will also reveal quiescent NS X-ray binaries 
and millisecond pulsars (some will be eclipsing), ideal targets to explore the upper bound of neutron 
star masses and the equation of state of ultra-dense matter (e.g. \citealt{linares18}).
For example, assuming a maximum NS mass of 2.3 M$_{\odot}$ \citep{ruiz18} 
the {\it photometric mass function} equation (eq. 2 in Paper III) implies that 
all NSs with FWHM>2200 \kms~must have orbital periods shorter than $\approx$3.6 h and will 
be mostly high inclination. Therefore, they can be easily spotted through eclipses, ellipsoidal modulation 
or irradiation variability in light curves spanning less than 4 hr.  

Beyond \ha~emitters, the colour-colour diagram will also reveal populations of AM CVn systems i.e. 
ultra-compact binaries with two (semi-)degenerate white dwarfs \citep{solheim10}.   
A combination of hydrogen deficiency, together with the presence of \ion{He}{I} emission lines, 
places these systems in a vertical stream at negative EWs. The region will also contain 
ultra-compact X-ray binaries with accreting NSs \citep{sengar17}. 
Ultra-compact binaries are very faint and rare, with a space density comparable to that of  
BHXBs, and thus only a few tens are currently known \citep{carter13}. Nonetheless, they are key systems 
to constrain common envelope parametrizations and, hence, close binary evolution models 
\citep{ivanova13}. Furthermore, they are predicted to be the brightest persistent GW sources and some 
will even become verification sources for LISA \citep{korol17,kupfer18}. 

Finally, because our narrow band filters are sensitive to gravity effects,  
$HAWKs$ will be able to discriminate between DA white dwarfs (WDs)
and A-B main sequence stars. WDs appear on a separate track under the main stellar locus, 
with DA2-3 types offset by as much as -0.1 mag in $\left(m_{r}-m_{H\alpha_{b}}\right)$ colour. 
With a survey depth of $r=24$ (5\%~photometry) $HAWKs$ can 
extend the number of DA WD discoveries to much larger volumes than previous surveys, including 
Gaia \citep{hollands18}. This will result in new constraints on their space density, scale height 
and merger rates \citep{giammichele12, kilic18}. The sample will likely contain new pulsating (ZZ Ceti) stars, 
WD binaries and WDs with planetary debris discs \citep{limoges15,farihi16}, which would be disclosed by 
follow-up studies.      

Synergies with existing and next generation surveys will be important to further disentangle and characterize 
these populations. For example, late-type T Tauri and coronal stars can be displaced rightward 
into the BHXB region due to the confluence of broad molecular bands and narrow \ha~emission. However, 
these are nearby objects which will be flagged by Gaia parallaxes and near-IR excesses. 
Contaminating symbiotic stars, on the other hand, can be identified from mid-IR colours granted by 
the extended all-sky survey NEOWISE  \citep{mainzer14}. 
Spitzer/GLIMPSE \citep{churchwell09} and NEOWISE mid-IR colours will also single out YSOs of different types 
through the presence of circumstellar gas/dust emission. 
Global photometric surveys such as Pan-STARRS \citep{kaiser10}, BlackGEM \citep{roelfsema16} and the 
{\it Large Synoptic Survey Telescope} LSST \citep{abell09} will feed in broad optical colours 
and variability information to help constrain spectral energy distributions (SEDs) and orbital periods. The 
ground-breaking sensitivity of the {\it Square Kilometer Array} SKA \citep{carilli04} and pathfinders, 
the {\it James Webb Space Telescope} JWST \citep{gardner06} and 
the {\it extended ROentgen Survey with an Imaging Telescope Array} eROSITA \citep{merloni12} will 
provide transient properties and multiwavelength fluxes to build full SEDs of all targets. 
Gaia will complement with distances and, therefore, luminosities for the brightest of these objects. 
Meanwhile, wide-field multi-object spectrographs such as $2dF$, WEAVE \citep{dalton12} or 4MOST \citep{dejong16} 
will furnish optical classification spectra of subsets of \ha~emitters in different FWHM bands. 
Finally, next generation ELTs will allow detailed spectroscopic studies of selected targets at the faint magnitude end 
and dynamical confirmation of new BH candidates.

\section*{Acknowledgements}

Based on observations made with the GTC and WHT telescopes under Director's Discretionary Time 
GTC/WHT/2017-089 of Spain's Instituto de Astrof\'isica de Canarias. 
We would like to thank T. Mu\~noz-Darias, P.A. Charles and T. Maccarone for many interesting discussions on the survey 
strategy and useful comments to the manuscript.  
We are very grateful to J. Calvo and the IAC mechanical engineers team for manufacturing a ring adaptor that 
allowed mounting filter MR661 in ACAM. Also to C. Benn and N. O'Mahony for their advice in the design of the adaptor ring and 
to T. Augusteijn, J. Telting and the NOT institute for the support and flexibility in providing us with the
\ha~filters \#21 and \#29. Observing support 
by ING support astronomers C. Fari\~na, R. Karjalainen, M. Karjalainen and L. Dom\'inguez is gratefully acknowledged. 
We also thank A. Cabrera and J. M\'endez for the flexibility and coordination in schedulling these DDT observations. 
We are grateful to J.M. Corral-Santana for providing us with spectral samples of \ha~emitting stars. 
JC acknowledges support by the Spanish Ministry of Economy, Industry and Competitiveness (MINECO) under grants 
EUIN2017-89095 and AYA2017-83216-P. MAPT also acknowledges support by MINECO under the Ram\'on y Cajal 
Fellowship RYC-2015-17854. 
MOLLY software developed by T. R. Marsh is gratefully acknowledged.







\begin{landscape}
 \begin{table}
	\centering
	\caption{Observing log}
	\label{tab:tab1}
	\begin{tabular}{lccccc}
		\hline
   Target  & Date & r & \hab  & \han & Spectra  \\ 
 		\hline
SWIFT J1357.2-0933 & 16 Feb 2018 & 30s & 300s & 600s & -- \\
GRO J0422+320 & 17 Feb 2018 & 2x20s/30s/2x60s  & 4x200s  & 100s/400s/2x200s  & 2x1200s  (OSIRIS) \\ 
A 0620-00            &  ,,  & 6x5s  & 5x30s  & 5x30s  &  12x60s (OSIRIS) \\
XTE J1118+480   &  ,,  & 2x5s/14x30s  & 2x30s/14x60s  & 30s/2x60s/10x120s/2x200s  & 8x300s (OSIRIS) \\
XTE J1859+226  & 20 June 2018 & 4x60s  & 3x270s/250s  & 206s/420s/3x600s  & 2x1800s  (OSIRIS) \\ 
		\hline
BD+30 2355 (A0 V) & 17 Feb 2018 & 0.2s & 5s & 10s & 30s  (ACAM) \\
FEIGE 15 (A1 IV) & ,, & 0.2s & 5s & 10s & 30s (ACAM) \\
SA95 15 (G8 V)  & ,, &  1s  &  8s & 30s & 150s  (ACAM) \\
SA95 16 (K7 V) & ,, & 2s & 8s & 30s & 150s (ACAM) \\
\hline
	\end{tabular}
\end{table}
\end{landscape}

\begin{landscape}
 \begin{table}
	\centering
	\caption{Photometric versus synthetic \ha~colours of standard stars. Typical errors are at the millimag level.}
	\label{tab:tab2}
	\begin{tabular}{lrcccc}
		\hline
STAR & \multicolumn{3}{c}{PHOTOMETRIC COLOURS} & \multicolumn{2}{c}{SYNTHETIC COLOURS} \\
     & $m_{r}^{\dagger}$  & $m_{r}-m_{H\alpha_b}$  & $m_{H\alpha_b}-m_{H\alpha_n}$ & 
 $m_{r}-m_{H\alpha_b}$  & $m_{H\alpha_b}-m_{H\alpha_n}$    \\
 		\hline
BD+30 2355 (A0 V) & 9.54   & 0       &  0       & 0.10 & 0.03   \\
FEIGE 15 (A1 IV)    &  9.32  & -0.01 & -0.06 & 0.06 & -0.01   \\
A0620-C1 (G5 V)~$^{\dagger\dagger}$   & 15.46  & 0.13 & 0.13   & 0.12 & 0.10   \\
SA95 15 (G8 V)      & 9.83    & 0.12 & 0.07   & 0.14 & 0.10    \\
SA95 16 (K7 V)     & 12.46   & 0.20 & 0.11  & 0.19 & 0.13      \\
\hline
	\end{tabular}
	\\
$^{\dagger}${~These are instrumental magnitudes. Comparison with calibrated r-band magnitudes indicate $r=m_{r}+1.1$.}	
\\
$^{\dagger\dagger}${~Photometric magnitudes and colours are weighted averages over all filter cycles.}
\end{table}
\end{landscape}

 \begin{landscape}
 \begin{table}
	\centering
	\caption{Photometric/synthetic EW and FWHM values compared to those measured from near-simultaneous spectra. 
	Errorbars represent 1 $\sigma$ confidence intervals.}
	\label{tab:tab3}
	\begin{tabular}{lccccccccc}
		\hline
      Target  & $m_{r}$ & SNR & $r^{\dagger}$ & EW & EW$_{\rm ph}$ & EW$_{\rm syn}$ & FWHM & FWHM$_{\rm ph}$ & FWHM$_{\rm syn}$ \\ 
                   & & & & (\AA)  & (\AA) & (\AA) &  (km/s)  & (km/s) & (km/s) \\
		\hline
GRO J0422+320 & 19.8 & 50 & 21.3 & 291$\pm$21 & 312$\pm$12 & 290$\pm$5 &1595$\pm$18 & 1464$\pm$132 & 1496$\pm$20 \\ 
A 0620-00            & 16.0 & 165 & 17.2 & 65$\pm$4     & 80$\pm$1  & 70$\pm$1 & 1766$\pm$20  & 1919$\pm$66 & 1839$\pm$11 \\
XTE J1118+480   & 18.0 & 150 & 19.1 & 90$\pm$3     & 81$\pm$1  & 98$\pm$1 & 2435$\pm$42  & 2197$\pm$86 & 2353$\pm$15 \\
XTE J1859+226   & 20.6 & 35 & 21.9 & 100$\pm$4   & 124$\pm$8 & 111$\pm$1 & 2323$\pm$47  & 2548$\pm$327 & 2236$\pm$64 \\
SWIFT J1357.2-0933$^{\dagger\dagger}$ & 19.6 & 16 & 20.9 & 103$\pm$15 & 79$\pm$14  & -- & 4173$\pm$203 & 3548$\pm$1045 & -- \\
\hline
	\end{tabular}\\
	$^{\dagger}$~Continuum r-band magnitude, calibrated from the instrumental magnitude $m_{r}$ and corrected 
	for the contribution of the \ha~flux following $r=m_{r} + 1.1+2.5 \log \left( 1+ EW/W_{r} \right)$.\\
	$^{\dagger\dagger}$~The quoted values for EW and FWHM are obtained from 2014 spectra presented in \cite{mata15}.  
\end{table}
\end{landscape}

\bsp	
\label{lastpage}
\end{document}